# Decomposition of Time Series Data to Check Consistency between Fund Style and Actual Fund Composition of Mutual Funds

Jaydip Sen and Tamal Datta Chaudhuri
Department of Analytics and Information Technology
Calcutta Business School
Diamond Harbour Road, Bishnupur – 743503, West Bengal, INDIA
email: jaydip.sen@acm.org, and tamalc@calcuttabusinessschool.org

**Abstract:** We propose a novel approach for analysis of the composition of an equity mutual fund based on the time series decomposition of the price movements of the individual stocks of the fund. The proposed scheme can be applied to check whether the style proclaimed for a mutual fund actually matches with the fund composition. We have applied our proposed framework on eight well known mutual funds of varying styles in the Indian financial market to check the consistency between their fund style and actual fund composition, and have obtained extensive results from our experiments. A detailed analysis of the results has shown that while in majority of the cases the actual allocations of funds are consistent with the corresponding fund styles, there have been some notable deviations too.

## 1. INTRODUCTION

Households and corporates invest in mutual funds as they do not have proper understanding of the financial markets, do not have the necessary expertise and do not have large amount funds to construct a diversified portfolio or portfolios. Mutual funds bring with them the expertise of fund managers and market research, the ability to balance risk and return, pool funds of small individuals or corporates to create a large corpus to build a diversified portfolio and also design funds to match the requirement of the customers. Individual investors also remain satisfied that their risk appetite is met, while the risk is managed by experts. Given the wide array of funds available, they can choose between sectoral funds, diversified funds, balanced funds, pure equity funds, pure debt funds, thematic funds, blue chip funds, value funds and money market funds. Mutual funds allow investment by individuals in G Secs and Call Money Market, where they otherwise would not be able to invest on their own.

A mutual fund is a professionally managed type of collective investment scheme that pools money from many investors and invests it in stocks, bonds, short-term money market instruments and other securities. Financial researchers and analysts have proposed several interesting ways of analyzing the composition and performance of mutual funds [1 - 8]. In this paper, we present a novel approach to study the characteristics of the individual stocks in a mutual fund that enables one to check the consistency between the fund style and the actual fund composition. While our proposed methodology is applicable for any type of mutual fund, the focus of this paper is on mutual funds that deal with equity only. Companies differ in terms of size, product, management quality, P/E multiple, popularity, momentum, dividend payout etc. It has been our contention that some company share prices show seasonality, some company share prices have a strong trend component, while some company share price movements exhibit a strongly random movement [9 - 12]. The objective of our work is to take a sample of equity mutual funds of varying style, and check whether the style proclaimed, matches with the time series decomposition of the price movements of the individual stocks of the fund. For example, a long term or large cap or blue chip equity fund should have stocks whose prices have a strong trend component, whereas a small cap equity fund should have a strong random component. Individuals looking for returns in the long run would opt for long term equity funds. It is expected that the fund managers would choose such stocks that have a strong trend component. Individuals with greater risk appetite looking for short term returns would look at small cap funds. Hence, such funds should have stocks with a strong random component.

The rest of the paper is organized as follows. In Section 2, we clearly define the problem at hand. Section 3 presents some related work on mutual fund analysis in the literature. Section 4 provides a detailed framework of methodology that we have used to solve our problem. In Section 5, we present an example of time series decomposition of stock prices and its associated results. Section 6 describes the summary of the decomposition of stocks for eight mutual funds that we have studied in this work. It also provides a detailed analysis of the composition of each of the mutual funds based on the decomposition results of the sample of stocks. Finally, Section 7 concludes the paper.

## 2. PROBLEM STATEMENT

The primary objective of our work is to develop a framework for checking the consistency among the actual fund composition and the fund style of several well-known mutual funds in the Indian financial market. While our proposed framework is applicable for any types of mutual fund, the focus of our endeavor in this paper is on the mutual funds that deal with equity funds only. In our previous work, we have demonstrated that some companies share prices exhibit seasonality, some are dominated by strong trend components, while some others are characterized by the presence of strong random components [9 – 12]. In this work, we take a sample of equity mutual funds of varying style, and check whether the style proclaimed matches with the time series decomposition of the price movements of the individual stocks of the funds. For example, a long term or a large cap or a blue chip equity fund should ideally include stocks whose prices have a strong trend component. On the other hand, a small cap fund is expected to be consisting of stocks with strong random components in their price movements.

For our study, we have considered eight mutual funds with varying fund style and fund capitalization. For each of these mutual funds, take a representative sample of stocks. The time series of the price movements of these stocks are extensively studied with respect to the characteristics of their three components, i.e., trend, seasonality and randomness, for a period of eight years (2008 – 2015). For each mutual fund, we analyzed the relative strengths of the three time series components for all the stocks in the sample. Based on the overall analysis of the time series components of the stocks, we check the consistency among the fund style and fund composition of the mutual funds. To the best of our knowledge, our approach is an entirely novel one for checking the consistency between the fund style and portfolio composition of a mutual fund. In Section 4 of this paper, we have provided a detailed description of the methodology that we have followed in our work.

## 3. REALTED WORK

Analysis of portfolio composition and performance of mutual funds have attracted considerable interest in the community of the financial researchers and analysts. Carhart [1] demonstrates that common factors in stock returns and investment expenses almost completely explain persistence in equity mutual funds' mean and risk adjusted returns [1]. The study also shows that the persistence is mutual fund does not reflect superior stock-picking skill. Chevalier and Ellison present the conflict between the mutual fund investors and the mutual fund companies [2]. The authors argue that while the investors would always prefer the fund companies to use their judgement to maximize the risk-adjusted returns, the fund companies would like to maximize their values by taking actions which lead to increase in the inflow of investments. Using a semi-parametric model the authors have estimated the shape of the flow-performance relationship for a sample of growth and growth and income funds during the period 1982 – 1992. Cremers and Petajisto introduce a new measure for portfolio management that represents the share of portfolio holdings that differ from the benchmark index holding [3]. The authors also demonstrate quantitatively how the new measure can be used to predict the performance of a mutual fund. Daniel et al. propose benchmarks for portfolio performance using parameters such as market capitalization, book-to-market, and prior-year return characteristics of 125 passive portfolios [4]. Based on their proposed benchmarks, the authors derived two measures: *characteristic timing* and *characteristic selectivity*. While the first measure enables portfolio managers to successfully time their portfolio weightings on these characteristics, the second measure allows the fund managers to select stocks that outperform the average stock having the same characteristics. Fama and French study luck versus skill in actively managed equity mutual funds, assuming that active funds with positive 'alpha' are balanced with negative 'alpha' [5]. In a cross-section, the authors find that true 'alpha' in net returns is negative for most active funds. Kacperczyk et al. argue that mutual fund managers may sometimes decide to concentrate their holdings in industries on which they have more informational advantages, and study the relation between the industry concentration and the performance of actively managed mutual funds in the US financial market during the period 1984 – 1999 [6]. The results indicate that, in general, more concentrated funds perform better when risk and style differences are controlled. Wermers investigates whether the stocks experiencing high levels of herding show a significant price adjustment and whether any such price adjustment is temporary or permanent [7]. The study shows that there is a relationship between abnormal stock returns and the direction of herding in the stock. The stocks that are bought in herds are found to outperform the stocks that are sold in herds during the following six months. Zheng studies the fund selection ability of the aggregate mutual funds investors' portfolio and observes that investors in aggregate are able to make buying and selling decisions based on good assessment of short-term future performance [8]. The author also observes that the trading strategies indicates a "smart money" effect in which the aggregate newly invested money in equity funds is able to forecast short-term future fund performance – funds that receive more money subsequently perform significantly better than those that lose money.

In contrast to all the aforementioned studies which have attempted to address several issues related to the performance of mutual funds, our proposed scheme is for checking the consistency between the fund style and actual fund composition. To the best of our knowledge, our proposition is a novel approach that is based on time series decomposition of individual stocks in a mutual fund.

## 4. METHODOLOGY

In this section, we provide a brief outline of the methodology that we have followed in our work. We have used the R programming language [13] for data management, data analysis and presentation of results. R is an open source language with very rich libraries that is ideally suited for any work that requires large scale data processing and analysis. In our work, first we have chosen eight popular mutual funds from the Indian financial market [14]. For each of these eight mutual funds, we have noted its portfolio composition and portfolio characteristics. For portfolio composition, we have identified the top ten sectors in which the mutual fund allocation has been done and for portfolio characteristics, we have noted the fund style and the degree of capitalization. Now, based on the top holdings of each of the mutual funds, we have taken a sample of 10 – 15 stocks from the top ten sectors in which fund allocation has been made. We use time series analysis using R programming language for studying the structural constituents for each stock. For this purpose, for each stock, we use its daily closing index value in the National Stock Exchange (NSE) for the period January 2008 to December 2015. We compute the monthly average of the index values of each stock and store the monthly average values in a plain text *(.txt)* file. Accordingly, for each stock, create a plain text file containing 96 (record of 8 years, each year containing 12 monthly averages) monthly average index values.

After the plain text files containing the monthly average values of the stock prices for all the 96 months under study are created for each of the stocks, we use the *scan( )* function defined in R to read the plain text file contents into an R object. The resultant R object for each stock is now converted into a time series object using the *ts( )* function defined in the TTR package in the R programming environment. For understanding the behavior of the time series object, we decompose it into its three constituent components: (i) trend, (ii) seasonal, and (iii) random, using the *decompose( )* function defined in the TTR package. The results of decomposition provide us a deeper insight into the behavior of the time series object corresponding to a particular stock. We also plot the time series and all its three components for each stock so as to get a visual idea about the relative strengths of the components in each time series.

After we decompose the time series of all the stocks in the sample for a particular mutual fund, we study the summary characteristics of the stocks in the mutual fund. For this purpose, for each stock, we compute the relative percentages of the trend, seasonal and random components with respect of its aggregate price for all months of the period under our study. We compute the maximum, the minimum and the mean values of such percentages so as to get an idea about the contributions of the three components to the overall aggregate price of the stock. The summary of the decomposition results for the stocks in the mutual fund is then compared with the fund style and the capitalization of the fund to verify whether the fund style is consistent with the fund composition or some notable inconsistencies exist.

For our study, we have considered eight mutual funds with varying fund style and fund capitalization. These eight mutual funds are: (1) UTI Infrastructure Fund, (2) Axis Midcap Fund, (3) ICICI Prudential Value Discovery Fund, (4) ICICI Focused Bluechip Equity Fund, (5) UTI Long Term Equity Fund, (6) Reliance Small Cap Fund, (7) ICICI Prudential Infrastructure Fund, and (8) UTI Bluechip Flexicap Fund. For each mutual fund, we have chosen sample of 10 – 15 stocks from the top ten sectors of allocation of the fund, and studied the time series characteristics of those stock in detail so as to identify their behavior. Based on our understanding of the behavior of the stocks, we have made an attempt to check the consistency of the portfolio allocation in the mutual fund.

## 5.   TIME SERIES DECOMPOSITION RESULTS

We now present the methods that we have followed in decomposing the time series of each stock that we have analyzed in our study. As already mentioned in Section 4, we have taken the daily closing index value in the NSE for the period January 2008 to December 2015. For each stock, we have computed the monthly average of the stock prices and stored the monthly average values in a plain text *(.txt)* file. Hence, corresponding to every stock, its plain text file contains 96 records (12 monthly average values for 8 years under study). The plain text file is converted into an R object on being read using the *scan( )* function in R and the resultant R object is converted into a time series objects using the *ts( )* function defined in the TTR package in the R computing environment. For each stock, the time series object is decomposed into its components using the *decompose( )* function in R. The results of decomposition are graphically presented using the *plot( )* function in R. As an illustration, we provide the details of the time series decomposition results for the HDFC Bank stock which is found to be present in some of the mutual funds in our study. Figure 1 depicts the time series of the monthly average price of the HDFC Bank stock for the period January 2008 to December 2015. It is clearly evident that the time series has a consistent upward trend with minor random fluctuations. Figure 2 presents the results of decomposition of the time series of the monthly average values for the HDFC Bank stock.

Table 1 presents the numerical values of the HDFC Bank stock time series data and its three components for the period January 2008 to December 2015. It may be interesting to observe that the values of the trend and the random components are not available for the period January 2008 – June 2008 and also for the period July 2015 – December 2015. The *decompose( )* function defined in the TTR package of R uses a moving average method with a period of 12 months for computing the trend component of a time series. Hence, in order to compute the trend value for the month of January 2008, we need time series data from July 2007 to June 2008. However, since we have used time series data from January 2008 to December

2015, the first trend value that the *decompose( )* function could compute was for the month of July 2008, and the last month being June 2015.

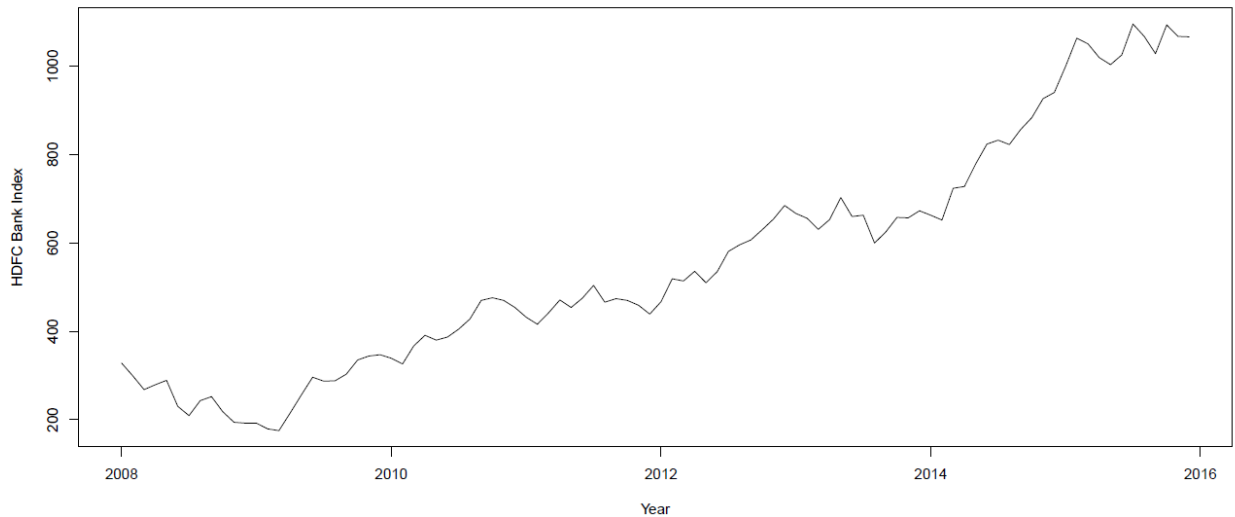

**Figure 1: HDFC Bank Stock Time Series (Period: January 2008 – December 2015)**

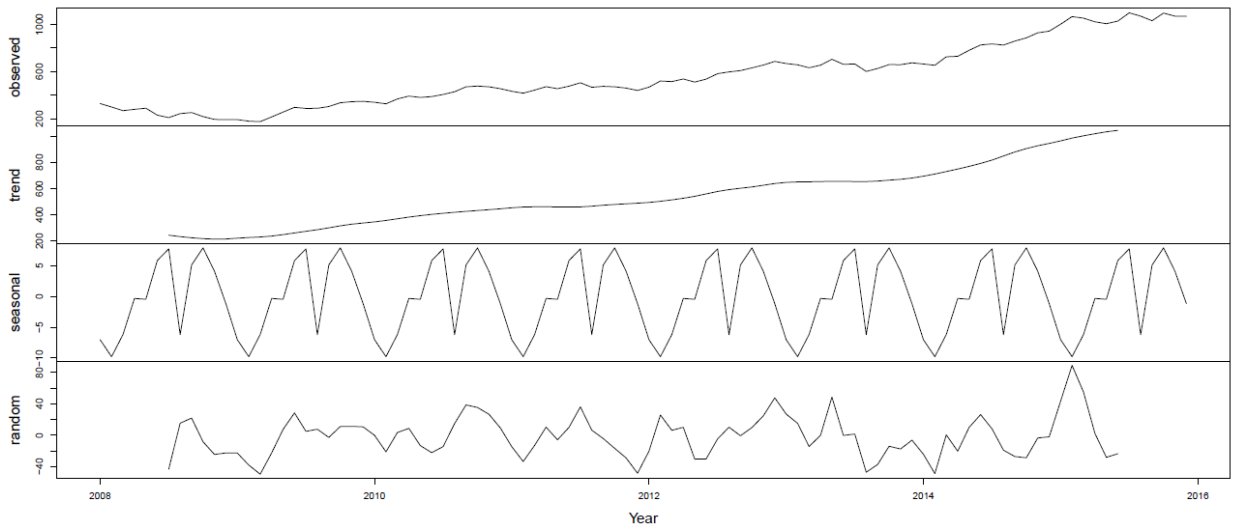

**Figure 2: Decomposition of the HDFC Bank Stock Time Series into its Components
(Period: January 2008 – December 2015)**

For computing the seasonal components, the *decompose( )* function first "detrends" (i.e., subtracts the trend components from the aggregate time series values) the time series, and then arranges the resultant new values in a 12-column format. Now, the seasonal value for each month is derived by computing the average of each column. It is easy to observe that the seasonal component for any particular month remains constant throughout the entire period under investigation. Finally, the random component of each month is obtained by subtracting the sum of the trend and the seasonal components from the aggregate time series value of the month. Since the trend values for the period January 2008 – June 2008 and July 2015 – December 2015 are not available, their random components could not be computed as well.

**Table 1: The Components of the Time Series of HDFC Bank Stock (Period: Jan 2008 – Dec 2015)**

| Year | Month | Aggregate | Trend | Seasonal | Random |
|---|---|---|---|---|---|
| 2008 | January | 328 | | -7 | |
| | February | 299 | | -10 | |
| | March | 268 | | -6 | |
| | April | 279 | | 0 | |
| | May | 289 | | 0 | |
| | June | 230 | | 6 | -43 |
| | July | 209 | 244 | 8 | 15 |
| | August | 243 | 234 | -6 | 22 |
| | September | 252 | 225 | 5 | -8 |
| | October | 218 | 218 | 8 | -24 |
| | November | 194 | 214 | 4 | -23 |
| | December | 192 | 216 | -1 | |
| 2009 | January | 192 | 222 | -7 | -23 |
| | February | 179 | 227 | -10 | -38 |
| | March | 175 | 231 | -6 | -50 |
| | April | 215 | 238 | 0 | -23 |
| | May | 256 | 249 | 0 | 8 |
| | June | 296 | 262 | 6 | 28 |
| | July | 287 | 274 | 8 | 5 |
| | August | 288 | 286 | -6 | 8 |
| | September | 303 | 301 | 5 | -3 |
| | October | 335 | 316 | 8 | 11 |
| | November | 344 | 328 | 4 | 11 |
| | December | 347 | 337 | -1 | 11 |
| 2010 | January | 339 | 346 | -7 | 0 |
| | February | 326 | 357 | -10 | -21 |
| | March | 367 | 370 | -6 | 4 |
| | April | 391 | 382 | 0 | 9 |
| | May | 380 | 394 | 0 | -13 |
| | June | 387 | 403 | 6 | -22 |
| | July | 405 | 412 | 8 | -14 |
| | August | 428 | 419 | -6 | 15 |
| | September | 470 | 426 | 5 | 39 |
| | October | 476 | 433 | 8 | 35 |
| | November | 470 | 439 | 4 | 27 |
| | December | 454 | 446 | -1 | 9 |
| 2011 | January | 432 | 454 | -7 | -15 |
| | February | 416 | 459 | -10 | -33 |
| | March | 442 | 461 | -6 | -13 |
| | April | 471 | 461 | 0 | 10 |
| | May | 454 | 460 | 0 | -6 |
| | June | 475 | 459 | 6 | 10 |
| | July | 504 | 460 | 8 | 36 |
| | August | 466 | 466 | -6 | 6 |
| | September | 474 | 473 | 5 | -4 |
| | October | 470 | 479 | 8 | -17 |
| | November | 459 | 484 | 4 | -29 |
| | December | 439 | 489 | -1 | -48 |
| 2012 | January | 467 | 494 | -7 | -20 |
| | February | 519 | 503 | -10 | 26 |
| | March | 514 | 514 | -6 | 6 |
| | April | 536 | 526 | 0 | 10 |

|      | Month     |      |      |     |     |
|------|-----------|------|------|-----|-----|
|      | May       | 510  | 541  | 0   | -30 |
|      | June      | 535  | 559  | 6   | -30 |
|      | July      | 581  | 578  | 8   | -5  |
|      | August    | 596  | 592  | -6  | 10  |
|      | September | 607  | 602  | 5   | -1  |
|      | October   | 630  | 612  | 8   | 10  |
|      | November  | 654  | 625  | 4   | 25  |
|      | December  | 685  | 638  | -1  | 48  |
| 2013 | January   | 667  | 647  | -7  | 27  |
|      | February  | 656  | 651  | -10 | 15  |
|      | March     | 631  | 652  | -6  | -14 |
|      | April     | 653  | 653  | 0   | 0   |
|      | May       | 703  | 655  | 0   | 49  |
|      | June      | 660  | 654  | 6   | 0   |
|      | July      | 663  | 654  | 8   | 2   |
|      | August    | 600  | 653  | -6  | -47 |
|      | September | 625  | 657  | 5   | -37 |
|      | October   | 658  | 664  | 8   | -14 |
|      | November  | 657  | 670  | 4   | -17 |
|      | December  | 673  | 680  | -1  | -6  |
| 2014 | January   | 663  | 694  | -7  | -24 |
|      | February  | 652  | 711  | -10 | -49 |
|      | March     | 724  | 730  | -6  | 1   |
|      | April     | 728  | 749  | 0   | -20 |
|      | May       | 779  | 769  | 0   | 10  |
|      | June      | 824  | 792  | 6   | 26  |
|      | July      | 833  | 817  | 8   | 8   |
|      | August    | 823  | 848  | -6  | -19 |
|      | September | 857  | 879  | 5   | -27 |
|      | October   | 884  | 905  | 8   | -29 |
|      | November  | 927  | 926  | 4   | -3  |
|      | December  | 941  | 944  | -1  | -2  |
| 2015 | January   | 1000 | 963  | -7  | 44  |
|      | February  | 1064 | 985  | -10 | 89  |
|      | March     | 1051 | 1002 | -6  | 55  |
|      | April     | 1020 | 1018 | 0   | 2   |
|      | May       | 1004 | 1033 | 0   | -28 |
|      | June      | 1026 | 1044 | 6   | -24 |
|      | July      | 1096 |      | 8   |     |
|      | August    | 1068 |      | -6  |     |
|      | September | 1029 |      | 5   |     |
|      | October   | 1094 |      | 8   |     |
|      | November  | 1068 |      | 4   |     |
|      | December  | 1067 |      | -1  |     |

## 6. ANALYSIS OF THE RESULTS

In this section, we present an analysis of the results that we have obtained from the time series decomposition of the sample of stock prices from the mutual funds under our study. For the purpose of understanding the relative strengths of the three components in a time series, we have followed a simple approach. For the trend component, we have computed three parameters: (i) maximum value of the percentage of the trend component with respect to the aggregate time series value, (ii) minimum value of the percentage of the trend component with respect to the aggregate time series value, (iii) mean value of the percentage of the trend component with respect to the aggregate time series value.

While the trend component in a time series is always positive, the same is not true for the seasonal and the random component. At certain points of time, a time series may have a negative seasonality and a negative random component, yet the time series may still exhibit a strong upward trend. At those points, the magnitude of the trend component of the time series will be greater than the aggregate value of the time series, since the aggregate value is the sum of the positive trend component and the sum of the two negative components (i.e., the seasonality and the random components). It is easy to understand that at those points of the time series, the percentage of the trend component with respect to the aggregate value of the time series will be greater than 100. We can observe from Table 2 – Table 9 that in all cases, the maximum percentage values of the trend components with respect to the overall aggregate values are greater than 100. For computing the mean percentage of the trend components, we consider the simple arithmetic average of the trend percentages values. It may be noted that the mean percentage of the trend components are computed using 84 observations since the trend percentage values are missing for 12 months (the first six months and the last six months of the period under study).

Unlike the trend component percentage values, the seasonal component percentage time series have values for all the 96 months of the period under our study. Hence, to maintain parity with the trend percentage time series values, we ignore the seasonality percentage values for the first six month and the last six months. Since seasonal component values are negative for certain months, the percentage values of the seasonal components with respect to the aggregate values of the time series for those months are also negative. As in the case of trend percentage values, we compute the maximum and the minimum values of the percentage of the seasonal components for every stock. However, for computing the mean of the percentage of the seasonal components, we consider the absolute values of the percentage records and ignore their signs. In other words, the mean value of the percentage of the seasonal components of the time series of a stock represents the average of their magnitudes only. This approach of computation of the mean percentage has been followed so that the mean truly reflects the overall seasonality percentage in the time series and the positive and the negative values cannot nullify each other.

The random component percentage values are computed in the same manner. Like the trend component percentage series, the random component series also has 84 records. Hence, unlike the seasonal component percentage series, we do not ignore any records for the random component percentage series. However, similar to the seasonal component percentage values, the random components percentages for certain months are negative. We follow the same approach for computing the maximum, the minimum and the mean values of the random component percentages as we have done for the seasonal components.

In the following, we present the result and the analysis of the time series decomposition of the stocks for the eight mutual funds that we have studied. We have considered a component as a "dominant" component in a time series if its mean percentage value exceeds the threshold value of 15.

**6.1 UTI Infrastructure Fund**

The investment objective of this fund is to provide income distribution and/or medium to long term capital appreciation by investing predominantly in the equity or equity related instruments in the companies engaged either directly or indirectly in the infrastructure growth of the Indian economy [15]. The fund style for this fund is 'blend' and the capitalization is 'medium'. The top ten sectors of allocation for this fund are: (i) construction, (ii) engineering, (iii) financial, (iv) energy, (v) consumer durable, (vi) services, (vii) diversified, (viii) automobile, (ix) communication, and (x) chemicals. Based on these ten sectors, we have chosen a sample of thirteen stocks in which the fund has its holdings. Table 2 presents the summary of the three components of the time series for these stocks. An infrastructure fund should ideally contain stocks which have dominant trend and random components. It can be observed from Table 2 that all the thirteen stocks have dominant trend component and seven of these stock have dominant random components as well. However, three stocks, viz. Voltas and Blue Star and Container Corporation were found to have significant seasonal components. Hence, we conclude that for the UTI Infrastructure Fund, the fund style and the capitalization of the fund has been consistent with it actual fund composition barring two notable exceptions - Voltas and Blue Star. Container Corporation is a company that provides logistic supports to its customers, and therefore, it is an important company for the infrastructure development in India. Hence, in spite of the presence of a dominant seasonal component in its stock prices, we do not consider its presence in an infrastructure fund as a deviation in the fund composition.

**Table 2: Summary Statistics for Stocks in the UTI Infrastructure Fund**

| Stock | Trend (T) | | | Seasonal (S) | | | Random (R) | | | Dominant Comp (s) |
|---|---|---|---|---|---|---|---|---|---|---|
| | Max | Min | Mean | Max | Min | Mean | Max | Min | Mean | |
| ABB | 151 | 83 | 103 | 10 | -10 | 4 | 16 | -45 | 28 | T + R |
| Adani Ports & Special | 146 | 75 | 102 | 5 | -15 | 3 | 20 | -31 | 17 | T + R |
| Axis Bank | 172 | 86 | 103 | 7 | -6 | 2 | 16 | -79 | 9 | T |
| Bharat Forge | 178 | 76 | 105 | 29 | -22 | 16 | 18 | -62 | 21 | T + R + S |
| Blue Star | 168 | 79 | 104 | 29 | -21 | 23 | 16 | -56 | 11 | T + S |
| Container Corporation | 123 | 88 | 101 | 25 | -13 | 22 | 21 | -19 | 18 | T + S + R |
| ICICI Bank | 164 | 86 | 103 | 5 | -8 | 2 | 12 | -59 | 8 | T |
| Kalpataru Power Trans. | 196 | 77 | 106 | 13 | -19 | 4 | 28 | -93 | 22 | T + R |
| Larsen & Toubro | 169 | 80 | 103 | 10 | -19 | 4 | 13 | -55 | 9 | T |
| Reliance Industries | 142 | 84 | 102 | 5 | -6 | 3 | 13 | -39 | 6 | T |
| State Bank of India | 140 | 84 | 102 | 4 | -10 | 2 | 14 | -31 | 8 | T |
| Ultratech Cement | 153 | 84 | 102 | 28 | -12 | 22 | 24 | -47 | 18 | T + S + R |
| Voltas | 218 | 82 | 106 | 31 | -37 | 25 | 59 | -89 | 13 | T + S |

**Table 3: Summary Statistics for Stocks in the ICICI Prudential Infrastructure Fund**

| Stock | Trend (T) | | | Seasonal (S) | | | Random (R) | | | Dominant Comp (s) |
|---|---|---|---|---|---|---|---|---|---|---|
| | Max | Min | Mean | Max | Min | Mean | Max | Min | Mean | |
| Axis Bank | 172 | 86 | 103 | 7 | -6 | 2 | 16 | -79 | 9 | T |
| CESC | 142 | 85 | 102 | 8 | -8 | 3 | 22 | -41 | 20 | T + R |
| Coal India | 119 | 84 | 100 | 9 | -7 | 3 | 24 | -14 | 20 | T + R |
| Container Corporation | 123 | 88 | 101 | 25 | -13 | 22 | 21 | -19 | 18 | T + S + R |
| FAG Bearings India | 131 | 85 | 103 | 13 | -16 | 2 | 15 | -30 | 18 | T + R |
| Grasim Industries | 160 | 83 | 102 | 6 | -4 | 2 | 19 | -56 | 8 | T |
| ICICI Bank | 164 | 86 | 103 | 5 | -8 | 2 | 12 | -59 | 8 | T |
| Kalpataru Power Trans. | 196 | 77 | 106 | 13 | -19 | 4 | 28 | -93 | 22 | T + R |
| Larsen and Toubro | 169 | 80 | 103 | 10 | -19 | 4 | 13 | -55 | 19 | T + R |
| ONGC | 130 | 82 | 101 | 7 | -8 | 4 | 32 | -22 | 21 | T + R |
| Power Grid Corp | 121 | 87 | 101 | 4 | -5 | 2 | 9 | -16 | 4 | T |

## 6.2 ICICI Prudential Infrastructure Fund

The objective of this open-ended equity scheme is to generate capital appreciation and income distribution to unitholders by investing in equity or equity related securities of the companies belonging to the infrastructure industries and balance in debt securities and money market instruments including call money[15]. The fund style for this fund is 'blend' and the capitalization is 'medium. The top ten sectors of allocation for this fund are: (i) energy, (ii) construction, (iii) engineering, (iv) financial, (v) services, (vi) diversified, (vii) metals, (viii) communication, (ix) banking and financial services, and (x) oil and gas. Based on these ten sectors, we have chosen a sample of eleven stocks in which the fund has its holdings. Table 3 presents the summary of the three components of the time series for these stocks. An infrastructure fund should ideally contain stocks which have dominant trend and random components. It can be observed from Table 3 that all the eleven stocks have dominant trend components and seven of these stocks have dominant random components as well. Only one stock – Container Corporation- exhibits the presence of a strong seasonal component. However, as explained earlier, in Section 6.1, we don't consider it as a deviation in the composition in an infrastructure fund. Hence, we conclude that for the ICICI Prudential Infrastructure Fund, the fund style and the capitalization of the fund has been consistent with it actual fund composition.

**Table 4: Summary Statistics for Stocks in the Axis Midcap Fund**

| Stock | Trend (T) | | | Seasonal (S) | | | Random (R) | | | Dominant Comp (s) |
|---|---|---|---|---|---|---|---|---|---|---|
| | Max | Min | Mean | Max | Min | Mean | Max | Min | Mean | |
| City Union Bank | 144 | 84 | 103 | 12 | -11 | 2 | 16 | -40 | 8 | T |
| CRISIL | 131 | 85 | 101 | 18 | -20 | 5 | 10 | -17 | 6 | T |
| Dish TV India | 175 | 80 | 104 | 9 | -17 | 3 | 21 | -75 | 10 | T |
| NIIT | 205 | 72 | 97 | 39 | -26 | 8 | 36 | -88 | 26 | T + R |
| Page Industries | 130 | 10 | 101 | 26 | -26 | 5 | 34 | -35 | 19 | T + R |
| Procter and Gamble | 123 | 87 | 101 | 7 | -9 | 1 | 13 | -23 | 6 | T |
| PVR | 146 | 81 | 104 | 34 | -28 | 9 | 23 | -80 | 11 | T |
| Sanofi India | 115 | 91 | 101 | 11 | -6 | 2 | 8 | -30 | 4 | T |
| Sundaram Finance | 139 | 78 | 102 | 17 | -10 | 3 | 24 | -31 | 8 | T |
| Torrent Power | 169 | 81 | 104 | 7 | -14 | 3 | 39 | -54 | 29 | T + R |

**Table 5: Summary Statistics for Stocks in the ICICI Prudential Value Discovery Fund**

| Stock | Trend (T) | | | Seasonal (S) | | | Random (R) | | | Dominant Comp (s) |
|---|---|---|---|---|---|---|---|---|---|---|
| | Max | Min | Mean | Max | Min | Mean | Max | Min | Mean | |
| Amara Raja Batteries | 194 | 85 | 105 | 79 | -48 | 9 | 24 | -147 | 25 | T + R |
| Ambuja Cement | 134 | 87 | 101 | 5 | -4 | 1 | 13 | -34 | 7 | T |
| Axis Bank | 172 | 86 | 103 | 7 | -6 | 2 | 16 | -79 | 9 | T |
| Bharat Forge | 178 | 76 | 105 | 29 | -22 | 16 | 18 | -62 | 21 | T + R + S |
| Bharti Airtel | 122 | 85 | 101 | 9 | -7 | 3 | 14 | -24 | 6 | T |
| Container Corporation | 123 | 88 | 101 | 25 | -13 | 22 | 21 | -19 | 18 | T + S + R |
| HDFC Bank | 132 | 89 | 101 | 4 | -6 | 1 | 9 | -19 | 5 | T |
| Hero Motocorp | 118 | 87 | 101 | 9 | -12 | 3 | 12 | -35 | 18 | T + R |
| ICICI Bank | 164 | 86 | 103 | 5 | -8 | 2 | 12 | -59 | 8 | T |
| Larsen and Toubro | 169 | 80 | 103 | 10 | -19 | 4 | 13 | -55 | 19 | T + R |
| Mahindra & Mahindra | 163 | 84 | 103 | 20 | -29 | 29 | 11 | -70 | 7 | T + S |
| State Bank of India | 140 | 84 | 102 | 4 | -10 | 2 | 14 | -31 | 8 | T |
| Tata Motors | 187 | 80 | 105 | 35 | -28 | 6 | 24 | -12 | 14 | T |

### 6.3 Axis Midcap Fund

The objective of this fund is to achieve long term capital appreciation by investing predominantly in equity and equity related instruments of mid-size companies, with the focus of the fund being to invest in relatively larger companies within this category [15]. The fund style for this fund is 'growth' and the capitalization is 'medium. The top ten sectors of allocation for this fund are: (i) financial, (ii) services, (iii) engineering, (iv) healthcare, (v) energy, (vi) chemicals, (vii) technology, (viii) FMCG, (ix) textiles, and (x) metals. Based on these ten sectors, we have chosen a sample of ten stocks in which the fund has its holdings. Table 4 presents the summary of the three components of the time series for these stocks. A mid cap fund should ideally consist of stocks which have market capitalization within INR 50 billion to INR. 200 billion [16]. They represent medium-size companies and investments in the mid cap stocks can bring higher returns in 3 to 5 years. Essentially a mid-cap fund should consist of stocks that have a dominant trend component along with a possible presence of a random component as well. It can be observed from Table 4 that all the ten stocks that we have studied under this fund have exhibited the presence of a strong trend component with four of them having a reasonably strong random component as well. Hence, we conclude that for the Axis Midcap Fund, the fund style and the capitalization of the fund has been consistent with it actual fund composition.

### 6.4 ICICI Prudential Value Discovery Fund

The objective of this open-ended diversified equity scheme is to provide long-term capital growth by investing primarily in a well-diversified portfolio of companies accumulated at a discount to its fair value [15]. The fund style for this fund is 'blend'' and the capitalization is 'large'. The top ten sectors of allocation for this fund are: (i) financial, (ii) energy, (iii) automobile, (iv) diversified, (v) construction, (vi) services, (vii) technology, (viii) engineering, (ix) chemicals, and (x) healthcare.  Based on these ten sectors, we have chosen a sample of thirteen stocks in which the fund has its holdings. Table 5 presents the summary of the three components of the time series for these stocks. The value discovery fund should ideally consist of stocks that have a strong long term trend component with a possible presence of random and seasonal component in the short term. It can be easily observed from Table 5 that all the thirteen stocks that we have analyzed under this fund have dominant trend component with six exhibiting the presence of an associated random component and three showing the presence of an associated seasonal component as well. Hence, we conclude that for the ICICI Prudential Value Discovery Fund, the fund style and the capitalization of the fund has been consistent with its actual fund composition.

### 6.5 ICICI Prudential Focused Bluechip Equity Fund

The objective of this open-ended equity fund is to generate long-term capital appreciation and income distribution to unitholders from a portfolio that is invested in equity and equity-related securities of companies belonging to the large cap domain [15]. The fund style is 'growth' and the capitalization is 'large'. The top ten sectors of allocation for this fund are: (i) financial, (ii) technology, (iii) energy, (iv) automobile, (v) healthcare, (vi) FMCG, (vii) diversified, (viii) communication, (ix) construction, and (x) metals. Based on these ten sectors, we have chosen a sample of sixteen stocks in which the fund has its holdings. Table 6 presents the summary of the three components of the time series for these stocks. The focused blue-chip fund should ideally consist of stocks of companies market leaders and typically have market capitalization in billions. The stocks in this category would exhibit strong upward trend components with negligible random and seasonal components. It can be observed from Table 6 that all the stocks that we have analyzed exhibited dominant trend components, with three stocks indicating the presence of a random component and one stock showing the presence of a seasonal component as well. Our previous work has revealed that the stocks of the companies in automobile sectors have the presence of a seasonal component and Mahindra and Mahindra is no exception [9,10]. The stocks of Coal India, Larsen and Toubro and Kotak Mahindra Bank have exhibited presence of reasonably strong random components which is not expected in blue-chip stocks. Hence, we conclude that for the ICICI Prudential Focused Bluechip Equity Fund, the fund style and the capitalization of the fund has been consistent with its actual fund composition, barring these four notable exceptions.

**Table 6: Summary Statistics for Stocks in the ICICI Prudential Focused Bluechip Equity Fund**

| Stock | Trend (T) | | | Seasonal (S) | | | Random (R) | | | Dominant Comp (s) |
|---|---|---|---|---|---|---|---|---|---|---|
| | Max | Min | Mean | Max | Min | Mean | Max | Min | Mean | |
| Axis Bank | 172 | 86 | 103 | 7 | -6 | 2 | 16 | -79 | 9 | T |
| Bharti Airtel | 122 | 85 | 101 | 9 | -7 | 3 | 14 | -24 | 6 | T |
| Coal India | 119 | 84 | 100 | 9 | -7 | 3 | 24 | -14 | 20 | T + R |
| Divi's Laboratories | 127 | 87 | 101 | 6 | -9 | 3 | 12 | -12 | 6 | T |
| Grasim Industries | 160 | 83 | 102 | 6 | -4 | 2 | 19 | -56 | 8 | T |
| HDFC Bank | 132 | 89 | 101 | 4 | -6 | 1 | 9 | -19 | 5 | T |
| ICICI Bank | 164 | 86 | 103 | 5 | -8 | 2 | 12 | -59 | 8 | T |
| Infosys | 124 | 89 | 101 | 13 | -12 | 4 | 14 | -31 | 6 | T |
| ITC | 113 | 91 | 101 | 6 | -6 | 2 | 10 | -17 | 3 | T |
| Kotak Mahindra Bank | 181 | 83 | 103 | 7 | -20 | 2 | 14 | -69 | 18 | T + R |
| Larsen and Toubro | 169 | 80 | 103 | 10 | -19 | 4 | 13 | -55 | 19 | T + R |
| Mahindra & Mahindra | 163 | 84 | 103 | 20 | -29 | 29 | 11 | -70 | 7 | T + S |
| Motherson Sumi Sys | 124 | 85 | 102 | 9 | -11 | 2 | 16 | -24 | 8 | T |
| Power Grid Corp. | 121 | 87 | 101 | 4 | -5 | 2 | 9 | -16 | 4 | T |
| Reliance Industries | 142 | 84 | 102 | 5 | -6 | 3 | 13 | -39 | 6 | T |
| Tata Motors | 187 | 80 | 105 | 35 | -28 | 6 | 24 | -12 | 14 | T |

**Table 7: Summary Statistics for Stocks in the UTI Long Term Equity Fund**

| Stock | Trend (T) | | | Seasonal (S) | | | Random (R) | | | Dominant Comp (s) |
|---|---|---|---|---|---|---|---|---|---|---|
| | Max | Min | Mean | Max | Min | Mean | Max | Min | Mean | |
| Axis Bank | 172 | 86 | 103 | 7 | -6 | 2 | 16 | -79 | 9 | T |
| Bharti Airtel | 122 | 85 | 101 | 9 | -7 | 3 | 14 | -24 | 6 | T |
| HDFC Bank | 132 | 89 | 101 | 4 | -6 | 1 | 9 | -19 | 5 | T |
| Hero Motocorp | 118 | 87 | 101 | 9 | -12 | 3 | 12 | -35 | 18 | T + R |
| ICICI Bank | 164 | 86 | 103 | 5 | -8 | 2 | 12 | -59 | 8 | T |
| Infosys | 124 | 89 | 101 | 13 | -12 | 4 | 14 | -31 | 6 | T |
| ITC | 113 | 91 | 101 | 6 | -6 | 2 | 10 | -17 | 3 | T |
| Larsen and Toubro | 169 | 80 | 103 | 10 | -19 | 4 | 13 | -55 | 19 | T + R |
| ONGC | 130 | 82 | 101 | 7 | -8 | 4 | 32 | -22 | 21 | T + R |
| Reliance Industries | 142 | 84 | 102 | 5 | -6 | 3 | 13 | -39 | 6 | T |
| State Bank of India | 140 | 84 | 102 | 4 | -10 | 2 | 14 | -31 | 8 | T |
| Sun Pharmaceuticals | 117 | 86 | 101 | 9 | -15 | 3 | 28 | -19 | 26 | T + R |
| TCS | 131 | 88 | 102 | 10 | -16 | 3 | 10 | -34 | 6 | T |

## 6.6 UTI Long Term Equity Fund

The objective of this fund is to invest in equities, fully convertible debentures and bonds and warrants of companies. It also invests in issues of partly convertible debentures and bonds including those issues on right basis subject to the condition that, as far as possible, the non-convertible portion of the debentures and bonds so acquired or subscribed are disinvested within a period of twelve months from their acquisition [15]. The fund style is 'growth' and the capitalization is 'large'. The top ten sectors of allocation for this fund are: (i) financial, (ii) technology, (iii) energy, (iv) healthcare, (v) services, (vi) engineering, (vii) construction, (viii) construction, (ix) FMCG, (x) automobile. Based on these sectors, we have chosen a sample of thirteen stocks in which the fund has its holdings. Table 7 presents the summary of the three components of the time series for these stocks. The long-term equity fund should ideally consist of stocks that are strongly dominated by their trend components with possible presence of mild random and seasonal components. In Table 7, it can be observed that all the thirteen stocks have exhibited strong trend components, with four of them showing the presence of an associated random component as well. Hence, we conclude that for the UTI Long Term Equity Fund, the fund style and the capitalization of the fund has been consistent with its actual fund composition.

**Table 8: Summary Statistics for Stocks in the Reliance Small Cap Fund**

| Stock | Trend (T) | | | Seasonal (S) | | | Random (R) | | | Dominant Comp (s) |
|---|---|---|---|---|---|---|---|---|---|---|
| | Max | Min | Mean | Max | Min | Mean | Max | Min | Mean | |
| Atul Industries | 151 | 77 | 78 | 86 | -126 | 12 | 102 | -10 | 42 | R + T |
| Chambal Fertilizers | 140 | 72 | 82 | 11 | -10 | 3 | 45 | -38 | 32 | R + T |
| Cyient Technology | 187 | 80 | 94 | 26 | -11 | 4 | 48 | -90 | 28 | R + T |
| Genus Power | 211 | 68 | 98 | 22 | -25 | 7 | 61 | -111 | 35 | R + T |
| GIC Housing Finance | 160 | 76 | 93 | 19 | 18 | 3 | 46 | -49 | 32 | R + T |
| HDFC Bank | 132 | 89 | 101 | 4 | -6 | 1 | 9 | -19 | 5 | T |
| Kalpataru Power Trans | 196 | 77 | 106 | 13 | -19 | 4 | 28 | -93 | 22 | T + R |
| Navin Fluorine Intl | 180 | 80 | 95 | 15 | -23 | 5 | 32 | -57 | 22 | R + T |
| NIIT | 205 | 72 | 97 | 39 | -26 | 8 | 36 | -88 | 26 | R + T |
| Radico Khaitan | 132 | 81 | 92 | 8 | -12 | 4 | 31 | -32 | 18 | R + T |
| VIP Industries | 170 | 73 | 97 | 120 | -75 | 10 | 96 | -140 | 34 | R + T |

### 6.7 Reliance Small Cap Fund

The primary objective of this scheme is to generate long-term capital appreciation by investing predominantly in equity and equity-related instruments of small cap companies and the secondary objective is to generate consistent returns by investing in debt and money related securities [15]. The fund style is 'growth' and the capitalization is 'small'. The top ten sectors of allocation for this fund are: (i) chemicals, (ii) construction, (iii) engineering, (iv) technology, (v) financial, (vi) FMCG, (vii) textiles, (viii) healthcare, (ix) services, and (x) communication. Based on these sectors, we have chosen a sample of eleven stocks in which the fund has its holdings. Table 8 presents the summary of the three components of the time series for these stocks. A small cap fund consists of stocks whose market capitalizations are low. These stocks are often looked upon as a platform to make big returns in short span of time. It is expected that most of the stocks in a small cap fund would be having a strong random component in their time series. It can be observed from Table 8 that ten among the eleven stocks that we have studied have exhibited the presence of a dominant random component. The only exception that we have found is the HDFC Bank stock, which is not essentially a large cap stock having a dominant trend component.  Hence, we conclude that for the Reliance Small Cap Fund, the fund style and the capitalization of the fund have been consistent with it actual fund composition with only one notable deviation being observed in the portfolio composition.

## 6.8 UTI Bluechip Flexicap Fund

The investment objective of this scheme is to achieve long-term capital appreciation and/or dividend distribution by investing in stocks that are "leaders" in their respective industries sectors/sub-sectors [15]. The fund style is 'growth' and the capitalization is 'large'. The top ten sectors of allocation are: (i) financial, (ii) healthcare, (iii) technology, (iv) FMCG, (v) automobile, (vi) engineering, (vii) services, (viii) chemicals, (ix) construction, and (x) metals. Based on these sectors, we have chosen a sample of eighteen stocks in which the fund has its holdings. Table 9 presents the summary of the three components of the time series for these stocks. The blue-chip flexicap fund should ideally consist of stocks that are strongly dominated by trend component with possible presence of mild random and seasonal components. It can be easily observed from Table 9 that all the eighteen stocks have exhibited strong trend component, with six of them showing the presence of an associated random component as well. Hence, we conclude that for the UTI Bluechip Flexicap Fund, the fund style and the capitalization of the fund has been consistent with its actual fund composition

**Table 9: Summary Statistics for Stocks in the UTI Bluechip Flexicap Fund**

| Stock | Trend (T) | | | Seasonal (S) | | | Random (R) | | | Dominant Comp (s) |
|---|---|---|---|---|---|---|---|---|---|---|
| | Max | Min | Mean | Max | Min | Mean | Max | Min | Mean | |
| Amara Raja Batteries | 194 | 85 | 105 | 79 | -48 | 9 | 24 | -147 | 25 | T + R |
| Axis Bank | 172 | 86 | 103 | 7 | -6 | 2 | 16 | -79 | 9 | T |
| Divi's Laboratories | 127 | 87 | 101 | 6 | -9 | 3 | 12 | -12 | 6 | T |
| eClerx Services | 167 | 87 | 105 | 40 | -31 | 5 | 29 | -106 | 22 | T + R |
| Havells India | 191 | 86 | 105 | 14 | -55 | 4 | 19 | -83 | 11 | T |
| HDFC Bank | 132 | 89 | 101 | 4 | -6 | 1 | 9 | -19 | 5 | T |
| Hero Motocorp | 118 | 87 | 101 | 9 | -12 | 3 | 12 | -25 | 18 | T + R |
| Hindustan Zinc | 148 | 81 | 102 | 10 | -9 | 2 | 18 | -44 | 8 | T |
| ICICI Bank | 164 | 86 | 103 | 5 | -8 | 2 | 12 | -59 | 8 | T |
| Infosys | 124 | 89 | 101 | 13 | -12 | 4 | 14 | -31 | 6 | T |
| ITC | 113 | 91 | 101 | 6 | -6 | 2 | 10 | -17 | 3 | T |
| Kotak Mahindra Bank | 181 | 83 | 103 | 7 | -20 | 2 | 14 | -69 | 18 | T + R |
| MindTree | 154 | 84 | 103 | 20 | -16 | 5 | 19 | -69 | 10 | T |
| Motherson Sumi Sys | 124 | 85 | 102 | 9 | -11 | 2 | 16 | -24 | 8 | T |
| Page Industries | 130 | 10 | 101 | 26 | -26 | 5 | 34 | -35 | 19 | T + R |
| Sun Pharmaceuticals | 117 | 86 | 101 | 9 | -15 | 3 | 28 | -19 | 26 | T + R |
| TCS | 131 | 88 | 102 | 10 | -16 | 3 | 10 | -34 | 6 | T |
| Torrent Pharma | 125 | 80 | 102 | 16 | -26 | 3 | 15 | -31 | 6 | T |

## 7. CONCLUSION

In this paper, we have presented a novel approach for checking the consistency between the style of a mutual fund and its actual fund composition. Our proposed scheme is based on time series decomposition of the individual stocks in a fund. Decomposition of the time series of the stock prices into its constituent components provides several useful insights that are suitably aggregated by several statistical computations to obtain an overall fund composition of a mutual fund. This fund composition is then compared with the fund style that was originally envisioned by the fund manager so as to check the consistency between the fund style and the actual fund composition. While our proposed scheme is absolutely generic and can be applied to mutual funds of any type, we have applied our framework on eight well known equity funds in the Indian financial market and obtained extensive results from our experiments. The analysis of the results have revealed that while in majority of the cases the actual allocations of funds are consistent with the corresponding fund style, there have been several notable deviations too. Moreover, the results obtained using this approach can be extremely useful for portfolio construction of stocks. By performing analysis on time series of several sectors and studying the behavior of their trend, seasonal and random components, portfolio managers and individual investors can very effectively take decisions about buy/sell of stocks and their appropriate timing. However, for speculative gains, the investors may target the sectors that exhibit the presence of dominant random components in the time series of their stock prices.

## REFERENCE

1. Carhart, M. (1997). On Persistence in Mutual Fund Performance. *Journal of Finance*, 52, 57-82.
2. Chevalier, J. and Ellison, G. (1997). Risk Taking by Mutual Funds as a Response to Incentives. *Journal of Political Economy*, 105, 1167-1200.
3. Cremers, M. and Petajisto, A. (2009). How Active is Your Fund Manager? A New Measure that Predicts Performance. *Review of Financial Studies*, 22(9), 3329-3365.
4. Daniel, K., Grinblatt, M., Titman, S., and Wermers, R. (1997). Measuring Mutual Fund Performance with Characteristic Based Benchmarks. *Journal of Finance*, Volume: 52, Issue: 3, pages:1035-1058.
5. Fama, E. F. and French, K. R. (2010). Luck versus Skill in the Cross-Section of Mutual Fund Returns. *Journal of Finance*, Volume: 65, Issue: 5, 1915-1947.
6. Kacperczyk, M., Sialm, C. and Zheng, L. (2005). On the Industry Concentration of Actively Managed Equity Mutual Fund. *Journal of Finance*, Volume: 60, Issue: 4, pages: 1983 - 2011.
7. Wermers, R. (1999). Mutual Fund Herding and the Impact on Stock Prices. (1999). *Journal of Finance*, Volume: 54, Issue: 2, pages: 581-622.
8. Zheng, L. (1999). Is Money Smart? A Study of Mutual Fund Investors' Fund Selection Ability. *Journal of Finance*, Volume: 54, Issue: 3, pages: 901-993.
9. Sen, J., and Datta Chaudhuri, T. (2016). Decomposition of Time Series Data of Stock Markets and its Implications for Prediction – An Application for the Indian Auto Sector. *Proceedings of the 2nd National Conference on Advances in Business Research and Management Practices (ABRMP'2016)*, Kolkata, India, January 8-9, 2016.
10. Sen, J., and Datta Chaudhuri, T. (2016). A Framework for Predictive Analysis of Stock Market Indices – A Study of the Indian Auto Sector. *Calcutta Business School (CBS) Journal of Management Practices*, Volume: 2, Issue: 2, pages: 1-20.
11. Sen, J. & Datta Chaudhuri T. (2016). An Alternative Framework for Time Series Decomposition and Forecasting and its Relevance for Portfolio Choice – A Comparative Study of the Indian Consumer Durable and Small Cap Sectors. *Journal of Economics Library*, Volume: 3, Issue: 2, pages: 303-326.
12. Sen, J., and Datta Chaudhuri, T. (2016). An Investigation of the Structural Characteristics of the Indian IT Sector and the Capital Goods Sector – An Application of the R Programming Language in Time Series Decomposition and Forecasting. *Journal of Insurance and Financial Management*, Volume: 1, Issue: 4, pages: 68-132.
13. Coghlan, A. (2015). A Little Book of R for Time Series. https://media.readthedocs.org/pdf/a-little-book-of-r-for-time-series/latest/a-little-book-of-r-for-time-series.pdf. (Accessed on October 2016).
14. Value Research. https://www.valueresearchonline.com (Accessed on October 2016)
15. Money Control. http://www.moneycontrol.com (Accessed on October 2016)
16. Equity Master. https://www.equitymaster.com/outlook/asset-allocation/Understanding-Equities.html (Accessed on October 2016)